\documentclass{moriond}

\def\Journal#1#2#3#4{{#1} {\bf #2}, #3 (#4)}

\def\NPB{{\em Nucl. Phys.} B}
\def\PLB{{\em Phys. Lett.}  B}
\def\PRL{\em Phys. Rev. Lett.}
\def\PRD{{\em Phys. Rev.} D}

\def\JPG{{\em J. Phys.} G}
\def\CPC{{\em Chin.Phys.} C}
\def\IJMPA{{\em Int. J. Mod. Phys.} A}
\def\be{\begin{equation}}
\def\ee{\end{equation}}
\def\bea{\begin{eqnarray}}
\def\eea{\end{eqnarray}}

\begin{document}
\vspace*{4cm}
\title{TWO-BODY CHARMED $B$ MESON DECAYS IN FACTORIZATION ASSISTED TOPOLOGICAL AMPLITUDE APPROACH}

\author{ CAI-DIAN L\"U AND SI-HONG ZHOU }

\address{Institute of High Energy Physics, Beijing 100049, People's Republic of China
         %2.~State Key Laboratory of Theoretical Physics, Institute of Theoretical Physics,\\
         %Chinese Academy of Sciences, Beijing 100190, People's Republic of China
         }

\maketitle
\abstracts{
We analyze the two-body charmed $B$ meson decays $B_{u,d,s} \to D^{(*)}P(V)$
in the factorization assisted topological amplitude approach, where $P(V)$
denoting a light pseudoscalar (vector) meson. Different from the conventional
topological diagram approach, flavor $SU(3)$ symmetry breaking effects
are taken into account. Therefore only four universal nonperturbative parameters
are introduced to describe the contribution from non-factorization diagrams
for all the decay channels. 
The number of free parameters and the $\chi^2$ per degree of freedom are both significantly reduced comparing with the conventional topological diagram approach.
With the 4 fitted parameters, we predict
the branching fractions of 120 decay modes induced by both $b\to c$ and $b\to u$
transitions, which are well consistent with the measured data or to be tested
on the future experiments. We also investigated the relative size of different topological diagrams, isospin violation,
flavor $SU(3)$ symmetry breaking effects, compared with previous approaches.}

\section{Introduction}\label{sec:1}

The charmed hadronic $B$ mesons decays $B\to D^{(*)} P(V)$ are of great interest
attributed to their theoretical application of heavy quark symmetry.   These processes
serve as a good testing ground for various theoretical issues in hadronic $B$ decays,
such as factorization hypothesis, flavor $SU(3)$ symmetry breaking, and isospin violation, which are  essential for the study  on $CP$ asymmetry in other channels. Experimentally, plenty of two-body charmed hadronic
$B$ decays have been observed from the heavy flavor experiments~\cite{Amhis:2014hma}.
In the theoretical side, the factorization of the color-favored decays has been proved
within the QCD factorization approach~\cite{Beneke:1999br}and the soft-collinear effective
theory~\cite{Bauer:2001cu}. However, the color suppressed modes was found with a very
large branching ratio experimentally, which provide evidence for a failure of the
naive factorization and for sizeable relative strong-interaction phases between
different isospin amplitudes~\cite{neubert}. This was confirmed in the perturbative
QCD (PQCD) approach based on $k_T$ factorization~\cite{Keum:2003js,Li:2008ts,Zou:2009zza}.
The rescattering effects of $B\to D^{(*)} P(V)$ had also been studied within some models~\cite{Chua:2007qw}. Under the assumption of the flavor $SU(3)$ symmetry, the global fits
were performed in the topological quark diagram approach~\cite{Chiang:2007bd}, where
the magnitudes and the strong phases of the topologically distinct amplitudes were studied,
but the information of SU(3) asymmetry was lost. Due to the large difference between
pseudoscalar and vector meson, their $\chi^2$ fit has to be performed for each category
of decays to result in three sets of parameters.

Recently, the factorization assisted topological amplitude (FAT) approach was proposed
to study the two-body hadronic decays of $D$ mesons~\cite{Li:2012cfa,Li:2013xsa}.
By involving the non-factorizable contributions and the SU(3) symmetry breaking effect,
most theoretical predictions of the $D$ decays were in much better agreement with experimental
data.  The prediction of direct CP asymmetry in D meson decays by this approach is in the best precision than before \cite{direct}.
In this framework, the two-body hadronic weak decay amplitudes are firstly decomposed
in terms of some distinct quark diagrams similar to the conventional topological diagrammatic
approach. Then in order to keep the flavor $SU(3)$ breaking effects in the decay amplitudes,
we factorize out the decay constants and form factors formally from each topological amplitude.
The topological amplitude is then universal for all decay channels after factorization of
those hadronic parameters that can be treated as nonperturbative parameters for
non-factorization topological diagrams or they are effective Wilson coefficients for
factorization contributions.

In the present work, we shall generalize the FAT approach to study the two-body charmed
non-leptonic $B$ mesons decays. Only 4 theoretical parameters need to be fitted from the
available experimental data for 31 decay channels.

\section{The Amplitudes of $B\to D^{(*)} P(V)$ decays in FAT Approach}\label{sec:2}

\begin{figure}[htb]
\begin{center}
\includegraphics[scale=0.25]{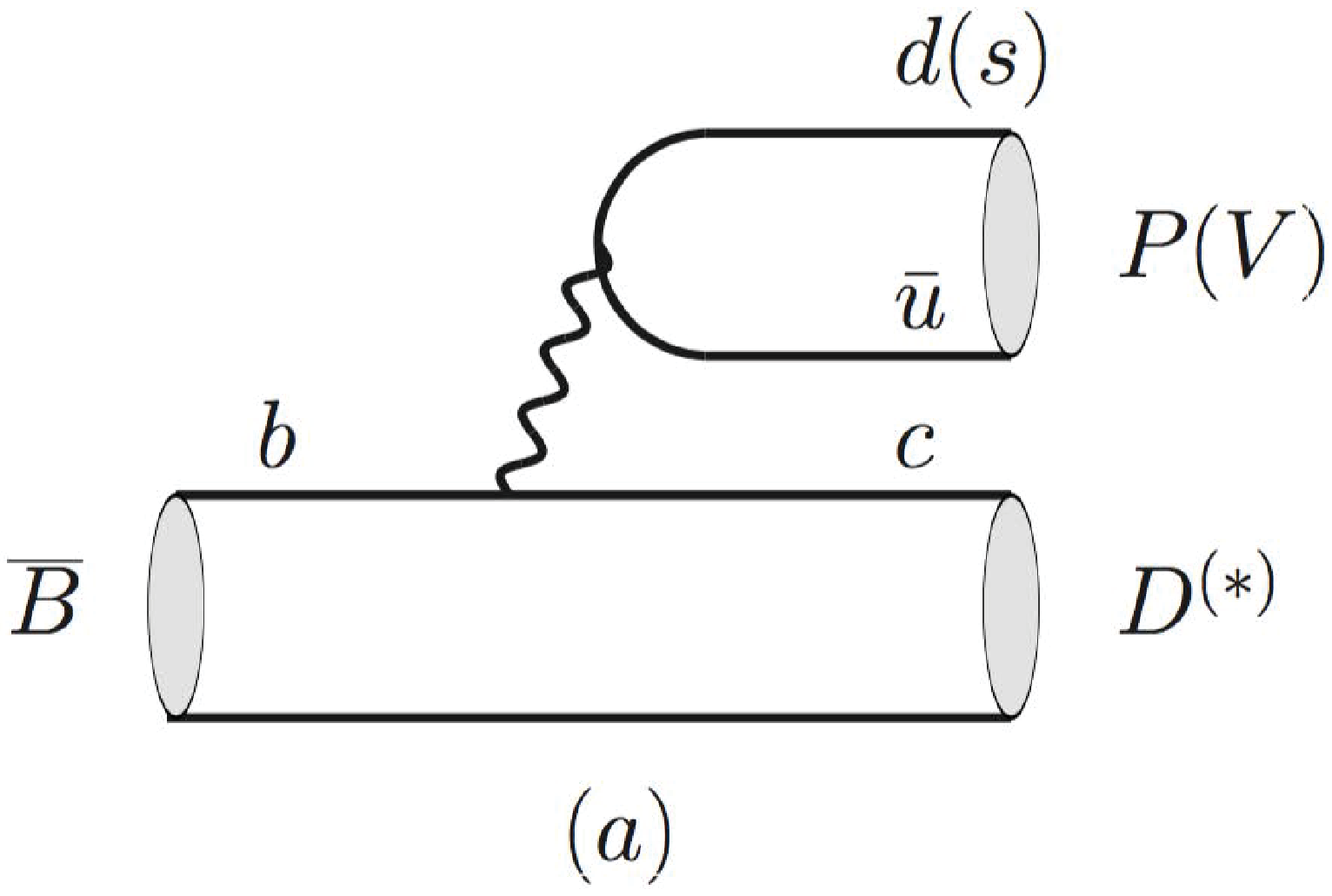}
\includegraphics[scale=0.25]{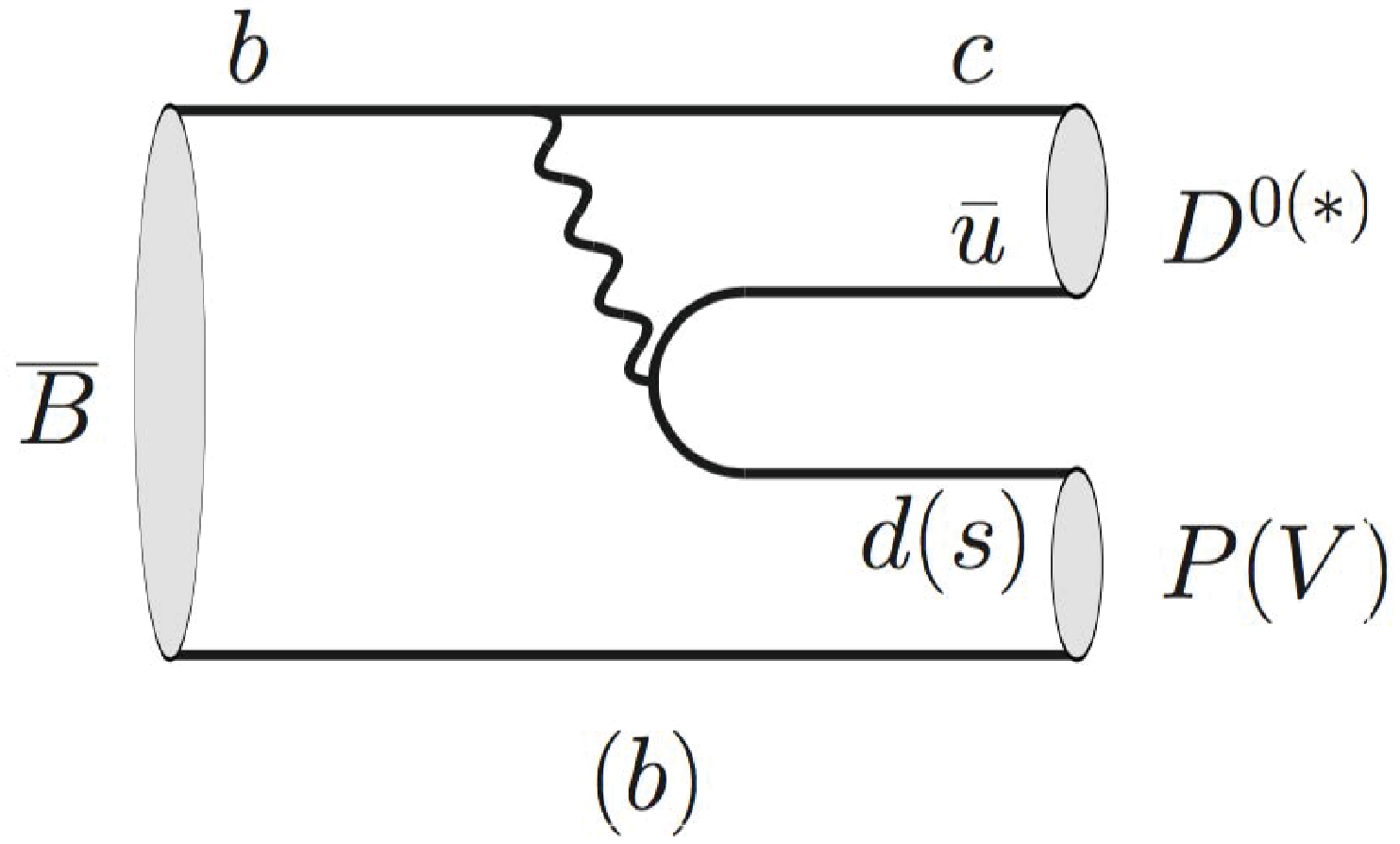}
\includegraphics[scale=0.25]{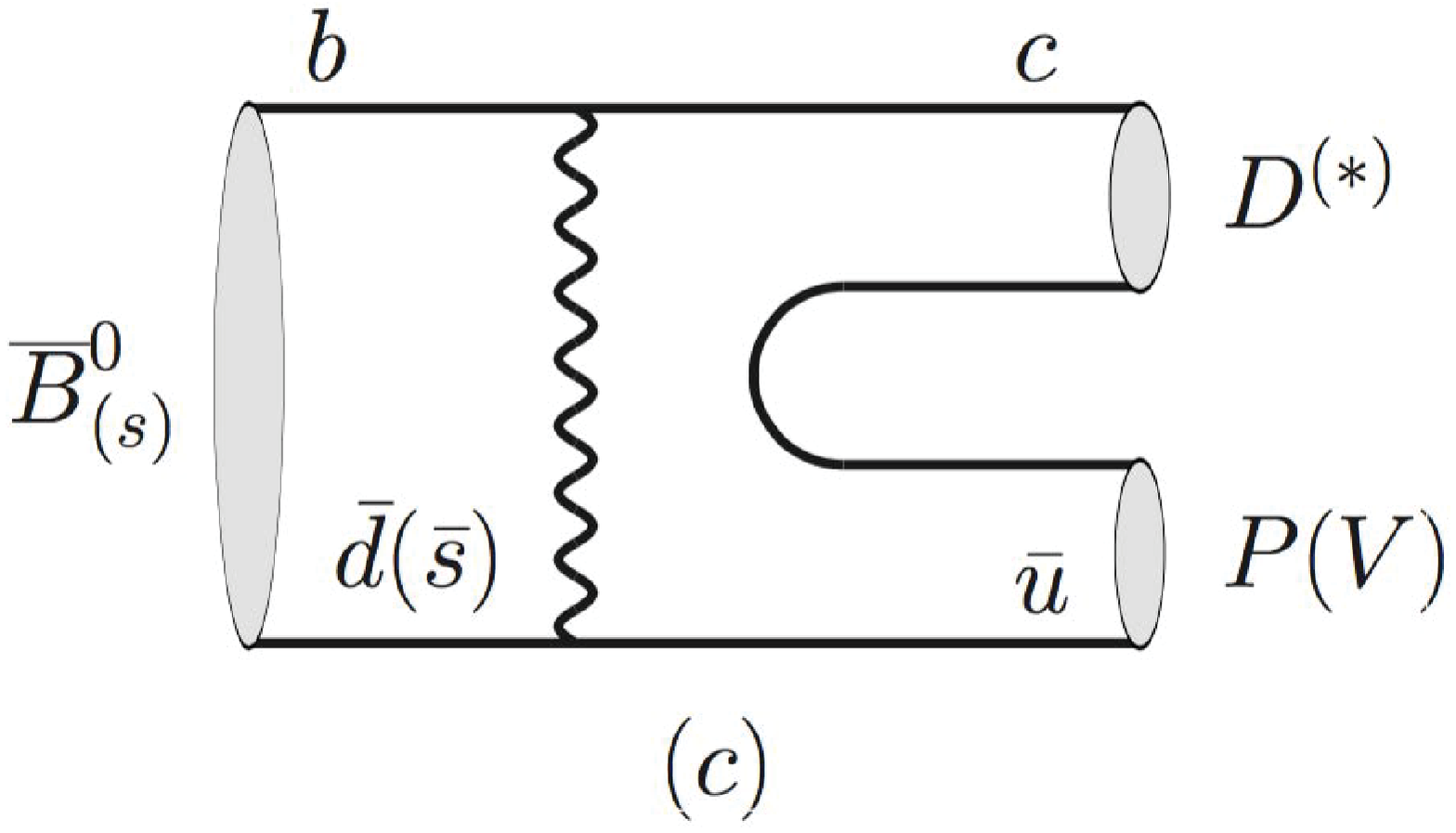}
\end{center}
\caption{Topological diagrams in the $b\to c$ transitions:
(a) the color-favored   tree diagram, $T$;
(b) the color-suppressed   tree diagram, $C$; and
(c) the $W$-exchange annihilation-type diagram, $E$.
Note that the $E$ diagram occurs only in the $\overline B_{d}^{0}$
and $\overline B_{s}^{0}$ decays. }
\label{fig:top}
\end{figure}

The topological diagrams in the $b\to c $ transitions includes color-favored tree
emission diagram $T$, color-suppressed tree emission $C$, and $W$-exchange diagram $E$,
as shown in Fig.\ref{fig:top}. Note that the $W$-annihilation diagram $A$ does not
occur in the $b\to c$ transition processes, but only appears in the $b\to u$ transitions.
In terms of the factorization hypothesis, the three diagrams of the
$\overline B \to D P$ modes can be written as
\begin{eqnarray}
T_c^{DP}&=&i{G_{F}\over\sqrt2}V_{cb}V_{uq}^{*}a_{1}(\mu)f_{P}
(m_{B}^{2}-m_{D}^{2})F_{0}^{B\to D}(m_{P}^{2}),\\
C_c^{DP}&=&i{G_{F}\over\sqrt2}V_{cb}V_{uq}^{*} f_{D}(m_{B}^{2}-m_{P}^{2})
F_{0}^{B\to P}(m_{D}^{2})\chi_c^{C}e^{i\phi_c^{C}},\\
E_c^{DP}&=&i{G_{F}\over\sqrt2}V_{cb}V_{uq}^{*}m_{B}^{2} f_{B}\frac{f_{D_{(s)}}
f_{P}}{f_{D}f_{\pi}} \chi_c^{E}e^{i\phi_c^{E}},
\end{eqnarray}
where the subscript $c$ stands for the processes induced by $b\to c$ transition.
$a_1$ is the effective Wilson coefficient for factorization diagram $T$. $f_{P}$
and $f_{D}$ are the decay constants of the light pseudoscalar meson and $D$ meson, respectively.
$F_{0}^{B\to D}$ and $F_{0}^{B\to P}$ are the scalar form factors of the $\overline B\to D$
and $\overline B\to P$ transitions. The contributions from non-factorization dominated
 diagram $C$ are parameterized as $\chi_c^{C}$, with its relative strong phase
$\phi_c^{C}$; while the contributions from W exchange  diagram $E$ are  $\chi_c^{E}$ and $\phi_c^E$.

Similarly to the amplitudes of $\overline B\to DP$ decays, the topological amplitudes of
$T$, $C$ and $E$ of the $\overline B\to D^{*}P$ and $\overline B\to DV$ decays can be
given respectively by
\begin{eqnarray}
T_c^{D^{*}P}&=\sqrt2{G_{F}}V_{cb}V_{uq}^{*}a_{1}(\mu)f_{P}m_{D^{*}}
A_{0}^{B\to D^{*}}(m_{P}^{2})(\varepsilon^{*}_{D^{*}}\cdot p_{B}),
\\
C_c^{D^{*}P}&=\sqrt2{G_{F}}V_{cb}V_{uq}^{*} f_{D^{*}}m_{D^{*}}F_{1}^{B\to P}
(m_{D^{*}}^{2})(\varepsilon^{*}_{D^{*}}\cdot p_{B})\chi_c^{C}e^{i\phi_c^{C}},\\
E_c^{D^{*}P}&=\sqrt2{G_{F}}V_{cb}V_{uq}^{*} m_{D^{*}}f_{B}\frac{ f_{D^{*}_{(s)}}
f_{P}}{f_{D}f_{\pi}}\chi_c^{E}e^{i\phi_c^{E}}  (\varepsilon^{*}_{D^{*}}\cdot p_{B});
\end{eqnarray}
and
\begin{eqnarray}
T_c^{DV}&=\sqrt2{G_{F}}V_{cb}V_{uq}^{*}a_{1}(\mu)f_{V}m_{V}F_{1}^{B\to D}(m_{V}^{2})
(\varepsilon^{*}_{V}\cdot p_{B}),
\\
C_c^{DV}&=\sqrt2{G_{F}}V_{cb}V_{uq}^{*} f_{D}m_{V}A_{0}^{B\to V}(m_{D}^{2})
(\varepsilon^{*}_{V}\cdot p_{B})\chi_c^{C}e^{i\phi_c^{C}} ,
\\
E_c^{DV}&=\sqrt2{G_{F}}V_{cb}V_{uq}^{*} m_{V} f_{B}\frac{f_{D_{(s)}}f_{V} }
{f_{D}f_{\pi}}\chi_c^{E}e^{i\phi_c^{E}}(\varepsilon^{*}_{V}\cdot p_{B}).
\end{eqnarray}
In above functions, $\varepsilon^{*}_{D^{*}}$ and $\varepsilon^{*}_{V}$
represent the polarization vectors of the  $D^{*}$ and $V$ meson, respectively.  $f_{D^{*}}$
and $f_V$ are the decay constants of the corresponding vector mesons.
$F_{1}^{B\to D}$ and $F_{1}^{B\to P}$ stand for the vector form factors
of $\overline B\to D$ and $\overline B\to P$ transitions, $A_{0}^{B\to D^{*}}$
and $A_{0}^{B\to V}$ are the transition form factors of $B\to D^{*}$ and
$\overline B\to V$, respectively. Note that, after factorizing out the corresponding
form factors and decay constants, we can use the same non-perturbative
universal parameters for all the three categories of $\overline B\to DP$, $\overline B\to D^{*}P$
and $\overline B\to DV$ decays. The total number of free parameters to be
fitted from experimental data is four. This is contrast to the
conventional topological diagram approach \cite{Chiang:2007bd},
where 15 parameters needed for the three categories of processes.

\section{Numerical results and discussion}\label{sec:3}

With the 31 experimental data induced by $b\to c$ transition \cite{PDG} and
using $\chi^2$ fit, we extract the four parameters with the best-fitted values
 as
\begin{eqnarray}
\chi_c^{C}=0.48\pm0.01,~~~
\phi_c^{C}=(56.6^{+3.2}_{-3.8})^{\circ},~~~
\chi_c^{E}=0.024^{+0.002}_{-0.001},~~~
\phi_c^{E}=(123.9^{+3.3}_{-2.2})^{\circ}, \label{fit}
\end{eqnarray}
with $\chi^{2}/d.o.f.=1.4$.
Even though with much more parameters than us, the $\chi^2$ per degree of
freedom is larger than ours in ref.~\cite{Chiang:2007bd}. With so many
parameters, they lost the predictive power of the branching fractions,
because there are not enough data of $B \to \overline D^{(*)}P(V)$ decays.
By contrast, with only 4 fitted parameters, we can predict 120 branching
fractions of $b\to c$ and $b\to u$ transition processes~\cite{Zhou:2015jba},
where we had employed an approximation that the four non-factorizable parameters
in the $b\to u$ processes are the same as those in the $b\to c$ processes.
The factorizable contribution from $W$ annihilation diagram $A$ are calculated in the pole
model~\cite{Kramer:1997yh}. Our results are consistent with experimental data
or to be tested in the LHCb and Belle-II experiments in the future.

The hierarchies of topological amplitudes are obtained as follows:
\begin{eqnarray}
|T_c^{DP}|:|C_c^{DP}|:|E_c^{DP}| &\sim & 1:0.45:0.1\\
|T_c^{D^*P}| :|C_c^{D^*P}| :|E_c^{D^*P}|&\sim & 1:0.36:0.1\\
|T_c^{DV}|:|C_c^{DV}|:|E_c^{DV}|&\sim &1:0.31:0.1.
\end{eqnarray}
It is obvious that the amplitudes of non-factorizable dominated color-suppressed
$C$ diagrams are relatively larger in the FAT approach compared with the
QCD-inspired methods~\cite{Beneke:1999br,Bauer:2001cu,Li:2008ts,Zou:2009zza},
for example, the relation were $|T_c^{DP}|\gg |C_c^{DP}| \sim |E_c^{DP}|$
in the PQCD approach. The relatively larger $C$ diagrams have significant
impacts on the processes without $T$ diagrams. For example, the topological
amplitudes of $\overline B^{0}\to D^{0}\rho^{0}$ and $D^{0}\omega$ decays
are $(E-C)/\sqrt2$ and $(E+C)/\sqrt2$, respectively. The branching fraction
of the $D^{0}\rho^{0}$ mode is predicted to be almost one half of that of
the $D^{0}\omega$ mode in the PQCD approach \cite{Li:2008ts}, since $C$ and
$E$ diagrams contribute destructively for the former mode but constructively
for the latter one, which does not agree with  the experiment. However,
this issue can be easily explained in the FAT approach in which both channels
are dominated by the $C$ diagram. It is easy to see that there is non-negligible difference for the $C$ contributions between different category of decays $B \to DP$, $B \to D^*P$ and $B \to D V$. This is the major reason that the conventional topological digram approach can not fit the three categories of decays together. 
On the other hand, the strong phase of $C$ diagram in eq.(\ref{fit}) is universal for all three kinds of decays, which agree with the soft-collinear effective theory \cite{scet}.

The isospin-amplitude ratio in $\overline B\to D \pi$ system showing significant
deviation from the heavy-quark limit can be traced back to the large color-suppressed
$C$ topologies due to ignored contributions from $E$ diagrams. The flavor $SU(3)$
symmetry breaking effect in $\overline B \to DM$  is about $10\sim20\%$ at the
amplitude level. By test the $SU(3)$ symmetry breaking effect in the
$B_u^-\to D^0 K^-$ and $B_u^-\to D^0\pi^-$, we conclude that the source of
$SU(3)$ symmetry breaking is mainly from the decay constants of light mesons
in $T$ diagram dominated decay modes as factorization hypothesis expected.
The $SU(3)$ symmetry breaking effect in $\overline B_s^0\to D_s^{*\mp} K^\pm$ and
$\overline B_s^0\to D_s^{*\mp} \pi^\pm$ is a little smaller than measurements implies
they might be more sizable than we expected.

\section{Conclusions}\label{sec:5}

Under the framework of the factorization assisted topological amplitude approach,
we analyzed $B\to D^{(*)} P(V)$ decays. By using the factorization results for $T$
diagram, only four universal
nonperturbative parameters  for non-factorization dominated $C$ and W exchange diagram $E$ were introduced to be  fitted from the 31 well measured
branching fractions. With the fitted results, we then predicted the branching
fractions of all 120 $B_{u,d,s}\to D^{(*)}P(V)$ decay modes. For the modes induced
by $b\to c$ transition, most results agree with the experimental data well. Comparing
with previous topological diagram analysis, the number of free parameters and the
$\chi^2$ per degree of freedom are both significantly reduced. The $SU(3)$ symmetry breaking is more
than $10\%$, and even reach $31\%$ at the amplitude level. The unmeasured branching
fractions, especially those processes dominated by $b\to u$ transition, and
possible larger $SU(3)$ symmetry breaking effects will be measured or tested in the ongoing LHCb experiment
and the forthcoming Bell-II experiment.

\section*{Acknowledgments}
We thank Y. B. Wei, Q. Qin, Y. Li and F. S. Yu for excellent collaborations. The work is partly supported by National Natural Science Foundation of China
(11375208, 11521505 and 11235005).

\end{document}